\documentclass{camera}

\newcommand{\AmS}{{\protect\the\textfont2
  A\kern-.1667em\lower.5ex\hbox{M}\kern-.125emS}}

\newcommand{\beq}{\begin{equation}}
\newcommand{\eeq}{\end{equation}}
\newcommand{\bea}{\begin{eqnarray}}
\newcommand{\eea}{\end{eqnarray}}

\def\dm2{\Delta m^2}
\def\sq2{sin^2(2\Theta)}

\input epsf

\begin{document}

\title{ICECUBE: THE STATE OF THE ART}

\author{TERESA MONTARULI for the IceCube Collaboration$^{(*)}$}

%
\organization{University of Wisconsin, Physics Deparment, Madison, WI, 53706, USA, \\ 
on leave of absence Universi\'a di, Bari, Dipartimento di Fisica, 70126, Italy\\
(*) http://icecube.wisc.edu/science/publications/vulcano2006.html\\
http://www.icecube.wisc.edu}

\maketitle

\begin{abstract}
In this paper we describe the performance of the 9 instrumented IceCube strings and 16 surface IceTop stations taking data at the Geographical South Pole after 2 deployment seasons.  We will focus on the description of the array and on the construction and data analysis status. 
The expected full array performance is discussed and compared to other results for the relevant physics studies using high energy neutrinos for astrophysical neutrino
searches.

\end{abstract}
\vspace{0.8cm}

\section{Status of high energy neutrino source searches}
\label{sec1}

A considerable effort is ongoing to broaden 
our knowledge of the universe through the observation of high energy neutrinos.
It is reasonable to believe, given also the increasing hints coming from gamma astronomy experiments, 
that the most powerful accelerators of the universe are sites of proton acceleration and interactions on
targets of matter and photons. Part of the energy could be transfered to neutrinos from meson decays
in astrophysical sources of high energy cosmic rays, such as Active Galactic Nuclei (AGNs) and GRBs
(Gamma-Ray Bursters), shock waves in SuperNova Remnants or micro-quasar binary systems exhibiting radio jets.
The operation of previous detectors and the underlying connections between
TeV-$\gamma$ production at sources indicate that $\nu$ fluxes are small and the detection of a positive signal may need years of operation of cubic-km scale detectors. 
Promising $\nu$ sources are those where strong $\gamma$ absorption is likely in the source, 
such as the micro-quasar systems LS 5029 and LSI +61 303 (which is below IceCube horizon), from which periodical TeV photon signals have been observed by HESS (Aharonian et al., 2006a)
and MAGIC (Albert et al., 2006), respectively. Whilst the $\gamma$-ray emissions from these sources 
can be weakened by electromagnetic cascading processes, neutrinos would
travel undisturbed to us producing up to a few tens of events in a km$^3$ detector per year (Aharonian
et al., 2006b). 

\begin{figure}[t!]
      \vspace{10truecm}
\includegraphics{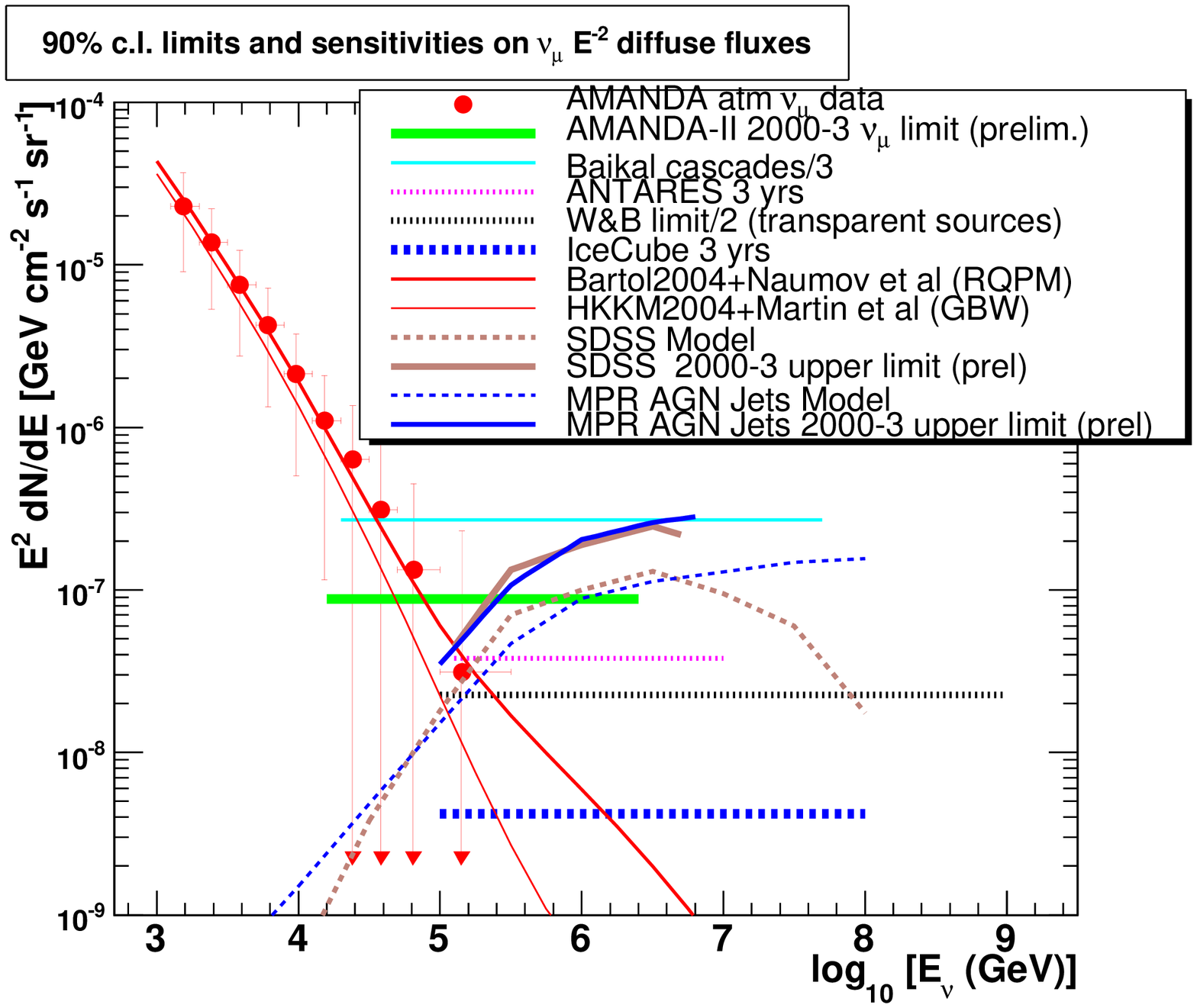}
\includegraphics{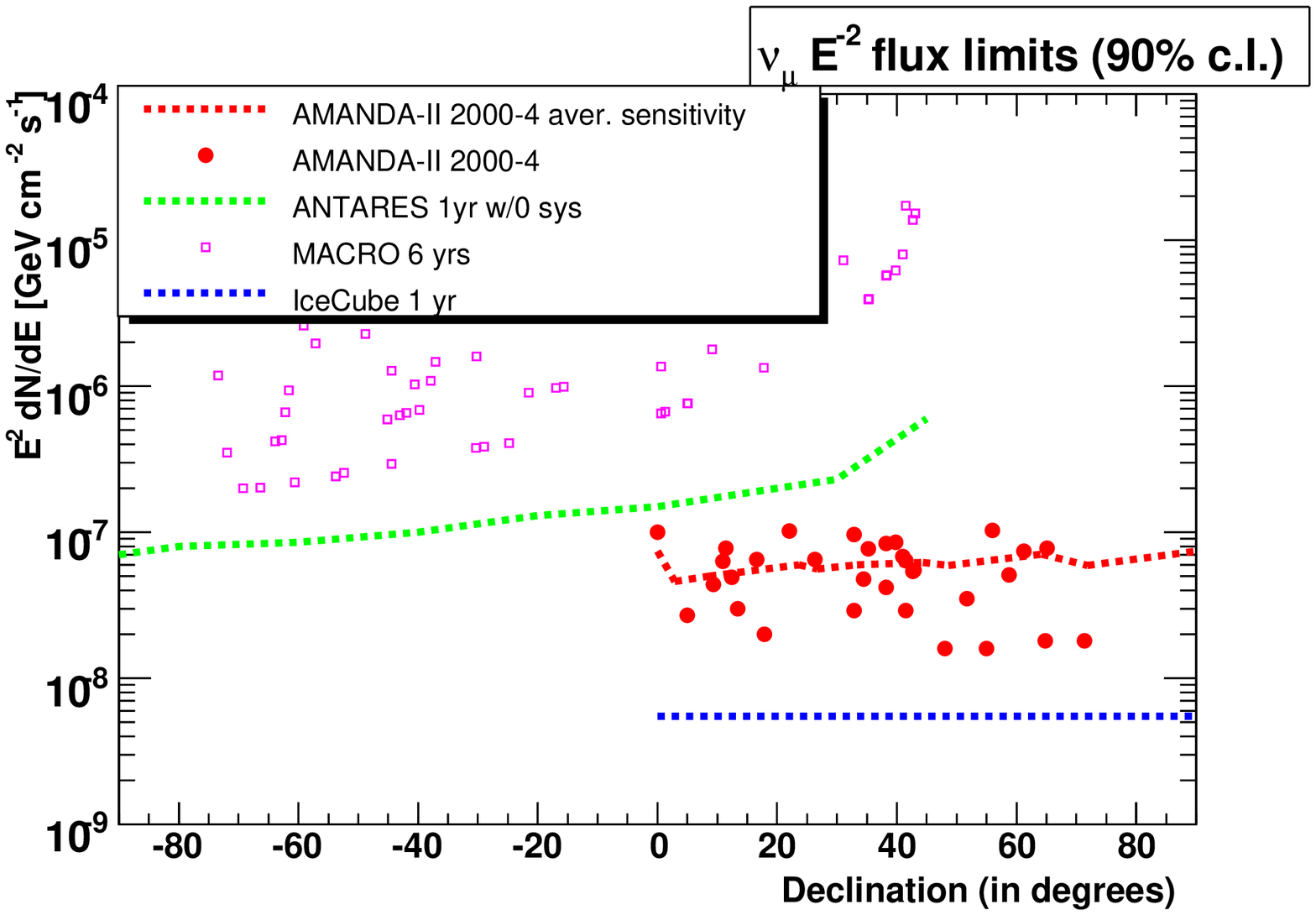}
 \caption[h]{{\bf On the left:}  The red full circles is the atmospheric $\nu$ flux measurement using
 2000 AMANDA-II data (Ackermann et al., 2005) with the central values for different  calculations shown as solid red lines (for a review see Honda and Gaisser, 2005).  
 The AMANDA-II upper limits for E$^{-2}$ $\nu_{\mu}$ diffuse fluxes (thick solid horizontal line) and for other models: limits = solid curved lines (G. Hill et al., 2006), models = dashed curved lines from AGN jet model in  (Mannheim et al, 2001) and AGN core model in (Stecker, 2005).
The Baikal limit for all flavor $\nu$s (upper line)  is shown  (Aynutdinov et al., 2006), divided by 3 to account for equi-partition of neutrino flavors due to oscillations.
 From bottom to top, the dotted horizontal lines are: sensitivity for 3~yrs of IceCube operation
 (Ahrens et al. 2004), Waxman \&Bahcall upper limit (Waxman \& Bahcall, 2000)  and 3~yrs ANTARES sensitivity (Montaruli et al., 2005).
  {\bf On the right:} upper limits on $E^{-2}$ $\nu_{\mu}$ fluxes 
 (90\% c.l.) vs declination in the Southern hemisphere from 1001 d of 2000-4 AMANDA-II data for selected sources (full red circles) and sensitivity (dashed line between circles)
 (Ackermann et al., 2006a). The IceCube sensitivity (lower dotted line) is from (Ahrens et al., 2004) for 1 yr. The MACRO upper limits (open squares) are from (Ambrosio et al., 2001). The ANTARES preliminary sensitivity for 1 full year of data is shown in the Northern
 hemisphere (Montaruli et al., 2005).}
     \label{fig1}
    \end{figure}

The state of the art of the detection of diffuse $\nu_{\mu}$ fluxes is summarized in Fig.~\ref{fig1}
(on the left). 
The Waxman \& Bahcall (WB) upper limit derived from measured $10^{18}$ eV cosmic rays (Waxman \& Bahcall , 2000) represents the maximum $\nu$ emission from optically thin extra-galactic sources. 
AMANDA-II recently presented the most stringent limit (G. Hill et al., 2006) and the Baikal 
experiment in a siberian lake is competitive in setting limits for diffuse fluxes using cascade-like
events (Aynutdinov et al., 2006). IceCube can observe fluxes well below this WB upper limit already in 1\,yr of active operation (Ahrens et al., 2004).
The most significant results for neutrino point-like source searches are currently produced by
the AMANDA-II detector: an average sensitivity for $E^{-2}$ neutrino fluxes of
$E^2 dN/dE \sim  6 \cdot 10^{-8}$ GeV cm$^{-2}$ s$^{-1}$ has been achieved  in 1001 days of operation during 2000-4.
In a sample of 4282 $\nu$ events (about 4/day) no evidence for a significant excess of $\nu$s
has been found. The largest fluctuation of the atmospheric $\nu$ background has a significance of 3.74$\sigma$, that occurs 69\% of the times on 100 randomized data samples. Upper limits for a few sources of interest are given in Tab.~\ref{tab1}.
 A search for the years 2000-4 is going to be published soon 
(Ackermann et al., 2006a).
IceCube capabilities are shown in Fig.~\ref{fig1} (on the right): the atmospheric $\nu$ event rate will be of the order of $10^5$/yr. The angular resolution (median of the angle between the $\nu$ and the reconstructed $\mu$ track)  will improve compared to the $2-3^{\circ}$ of AMANDA-II better than 
$\sim 1^{\circ}$. 
Notice that the results shown in Fig.~\ref{fig1} for IceCube from (Ahrens et al., 2004) are obtained using
AMANDA reconstruction adapted to the different geometry. We are currently working on a reconstruction based on the full knowledge of the pulse waveforms (see Sec.~\ref{sec2}) from which we expect a considerable improvement. Moreover, it should be noted that the sensitivities for IceCube shown in Fig.~\ref{fig1} include the contribution of a high prediction for the high energy charm contribution.
As discussed in (Ahrens et al., 2004) lower predictions will result in a less conservative estimate. 

\begin{table}[t!]
\begin{center}
\caption{\label{tab1}Upper limits to $E^{-2}$ neutrino fluxes for $E_{\nu} > 10$~GeV and measured and expected background events for a few of the monitored neutrino source candidates in 1001 days
of AMANDA-II live-time in 2000-4.}
\bigskip
\begin{tabular}{llll}
\hline
Source Name & N. obs. & N. exp. & Upper limit (90\% c.l.) \\
                          &              &                & $10^{-9}$ cm$^{-2}$ s$^{-1}$\\
\hline
Crab   & 10   & 6.74 & 10.2\\
Mrk 501   &  8              & 6.39 & 8.5\\
LSI +61 303   &  5             & 4.81 & 7.4\\
1ES 1959+650   &  5 &4.77              & 7.8\\
\hline
\end{tabular}
\end{center}
\end{table}

Neutrino telescopes are aimed at the detection of high energy $\nu$s ($> 100$ GeV)
through the Cherenkov light produced by $\nu$ induced charged products. 
Neutrinos of all flavors could reach us from sources, after long baseline oscillations 
transform the original $\nu_{\mu}$ and $\nu_{e}$ fluxes from meson decays in sources.
Muon induced $\nu$s point back to their sources for energies
$> 1$\,TeV, since the kinematical muon-neutrino angle becomes small, and the
long muon tracks allow precise reconstruction through the arrival times of detected Cherenkov photons.
This, together with the advantage of the target mass increase due to the 
muon range, makes 3 dimensional arrays of PMTs located in the depths of the ice or in deep sea
or lake water, telescopes capable at pointing back to sources with resolutions at degree level or better. 
Cascades induced by neutral current neutrino interactions or by tau and electron neutrino charged current interactions, though with much reduced pointing capabilities, allow for searches of diffuse fluxes thanks to the harder spectral properties
of neutrinos from shock acceleration mechanisms compared to neutrinos produced
in atmospheric cascades.
These atmospheric neutrinos represent from one side a background for astrophysical searches,
from the other a calibration tool and a signal for high energy interaction studies.
Similarly the atmospheric muons are a dangerous background, as well a tool for understanding the
detector with huge statistics. Their discrimination requires good reconstruction capabilities to single out
upward-going events that could be only induced by $\nu$s crossing the Earth, or 
potential to identify events with the interaction vertex in or close to the interaction region.

\section{IceCube}
\label{sec2}

The IceCube Collaboration, comprising about 250 members from 30 Institutions in USA, Europe, New Zealand and Japan, is building the first cubic kilometer detector in the Antarctic ice at depths between 1450 and 2450 m. 
 The Collaboration acquired the know-how for the detector running and data handling
 from the predecessor AMANDA experience. The AMANDA-II detector, taking data since 1996 and completed in 2000, is made of 677 optical modules (pressure resistant glass spheres) containing 
 $20$\,cm photomultipliers (PMTs) on 19 strings arranged in a cylindrical geometry with 200 m diameter at a depth between 1500-1950 m. The PMT signals are sent to the surface over either twisted pair, coaxial cables or optical fibers and are processed by the surface data acquisition system. The trigger requires the coincidence of  24 hit PMTs in 2$\mu$s and the trigger rate is around 80-90 Hz.
 The operation of this detector has demonstrated that deep ice is a suitable medium
for astronomy, characterized by an effective scattering length of about  20\,m 
and an absorption length of about 110~m at 440\,nm. 

\begin{figure}[t!]
      \vspace{10truecm}
\includegraphics{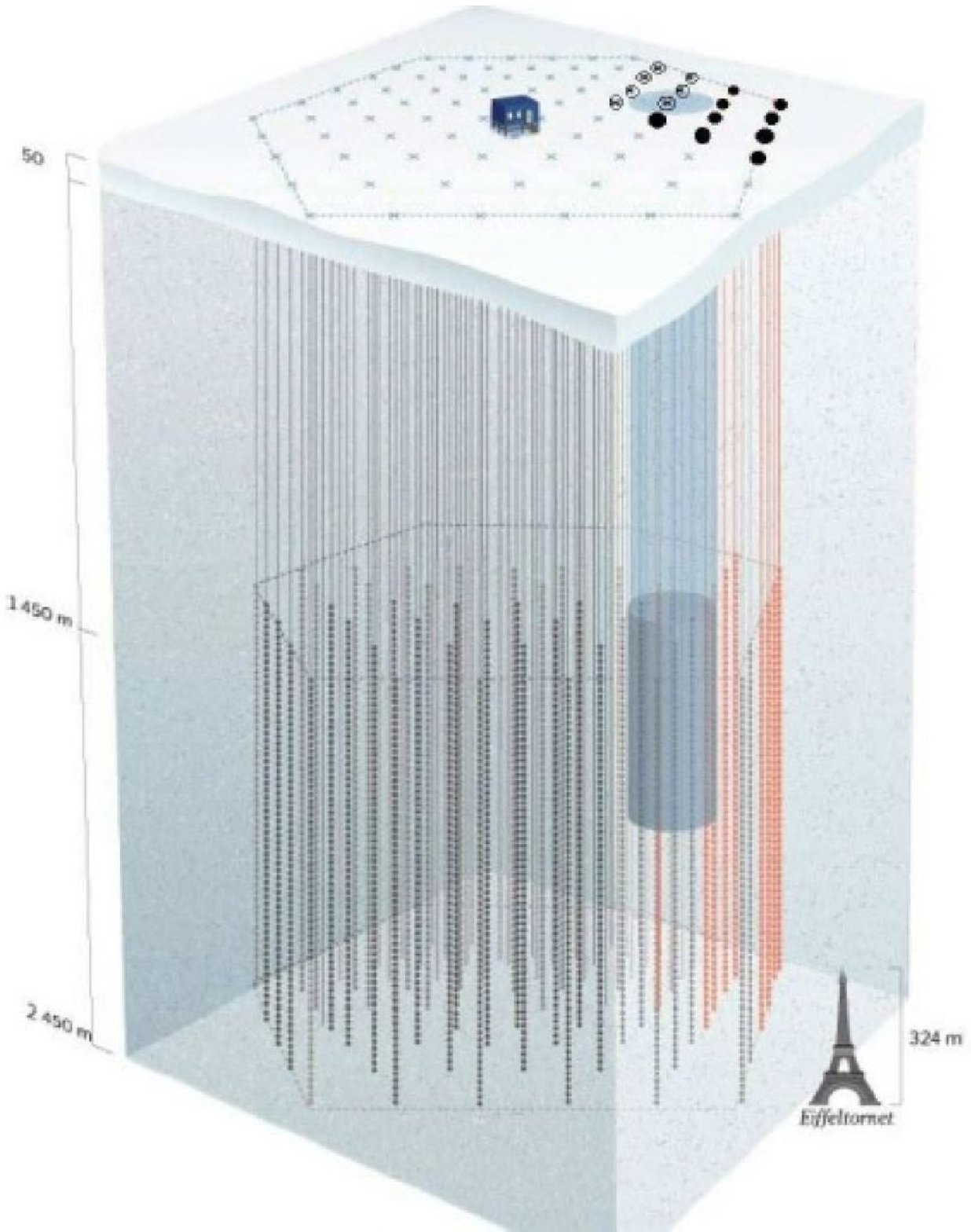}
\includegraphics{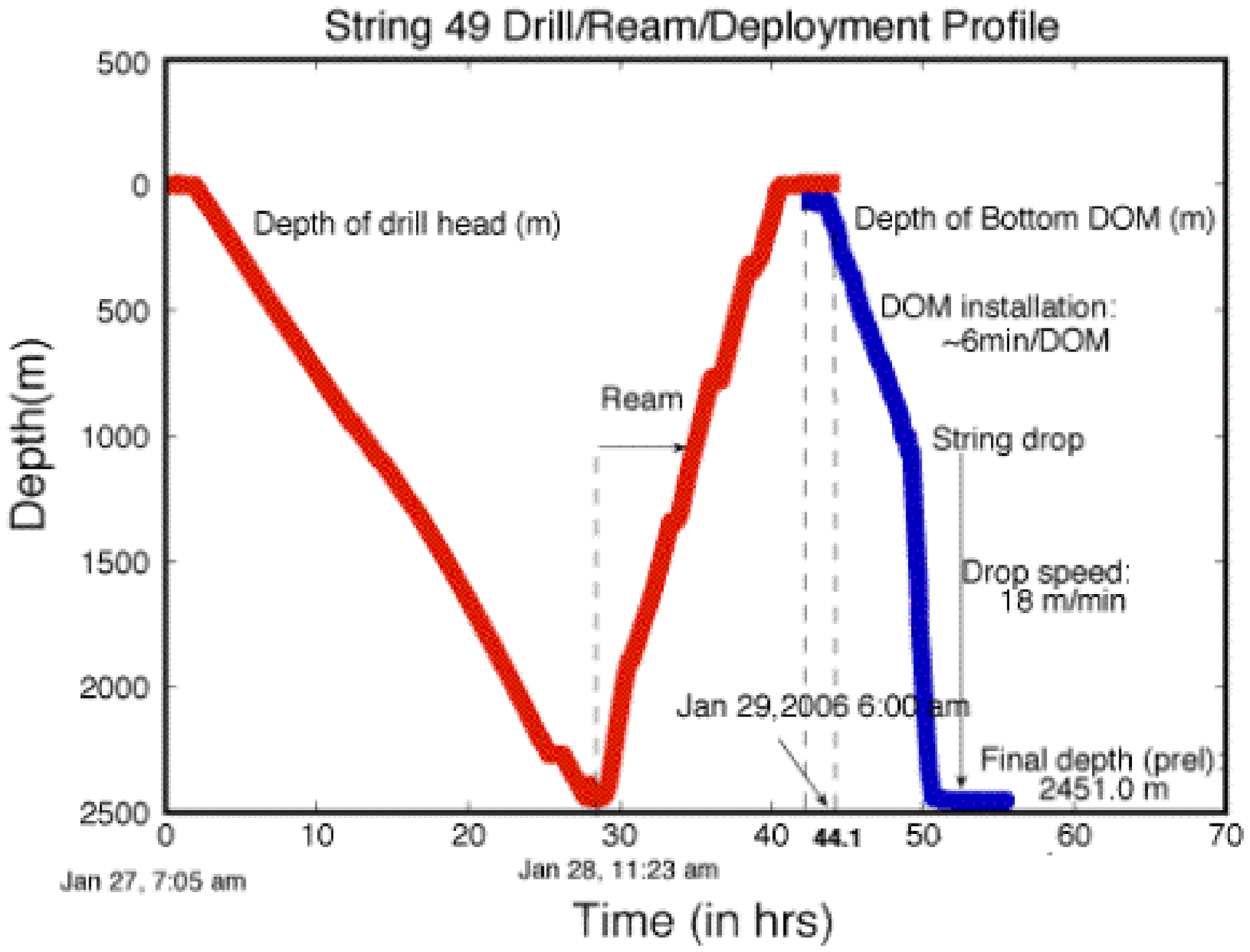}
      \caption[h]{{\bf On the left:} Pictorial view of the IceCube and AMANDA (darker small cylinder) deep ice detectors and of the IceTop surface array. The full circles correspond to the completed IceTop stations and to the positions of the IceCube strings. {\bf On the right:} depth vs time employed to install one of the 8 strings during last season (30 hrs for the drill head descent to 2450 m, 10 hrs for the ream and further 6 min/DOM for the installation of 60 DOMs).}
     \label{fig2}
    \end{figure}

 The construction goal of the IceCube Collaboration is to install 80 strings 
 between depths of 1450 m to 2450 m, each with 60 Digital Optical Modules (DOMs) spaced by 17 m. The distance between strings, arranged in an hexagonal pattern, 
is about 125 m. AMANDA, with inter-string distances between 30-50\,m, will be enclosed in the IceCube array and will serve as a denser sub-detector capable of detecting lower energy events (see Fig.~\ref{fig2} on the left). The deep ice detector will be complemented by a surface array, IceTop, with
stations of 2 tanks of frozen water seen by 2 DOMs at 10~m distance one from the other corresponding to string holes. 
This extensive air shower array will be used for purposes of angular resolution
calibration and atmospheric muon veto and for studying the cosmic ray composition in the
sub-PeV-EeV primary energy region.

During 2004-2005 austral summer, the first IceCube string (String 21) and 4 IceTop stations were installed and commissioned, demonstrating the proper operation of the hot-water drilling system, DOM electronics, data acquisition, calibrations and slow-control (Ackermann et al., 2006b). 
During the 2005-6 season, further 8 strings were deployed and 24 IceTop tanks. The survival rate of
the 604 frozen DOMs is about 98\%. Preliminary analyses show that the trigger rate 
(for a request of at least 8 DOMs hit with nearest-neighbour local coincidences in 5$\mu$s) is $\sim 140 $\,Hz, in agreement with Monte Carlo expectations.
The drilling system developed for IceCube proved a tremendous improvement compared to the 
one used for AMANDA: while for AMANDA it took about 90 hrs to reach a maximum depth of 1950 m and further 18 hrs for the deployment, IceCube drilling takes $\sim 40$\,hrs reaching 2450 m and the deployment takes 10 hrs (see Fig. ~\ref{fig2} on the right). 

\begin{figure}[t!]
      \vspace{10truecm}
\includegraphics{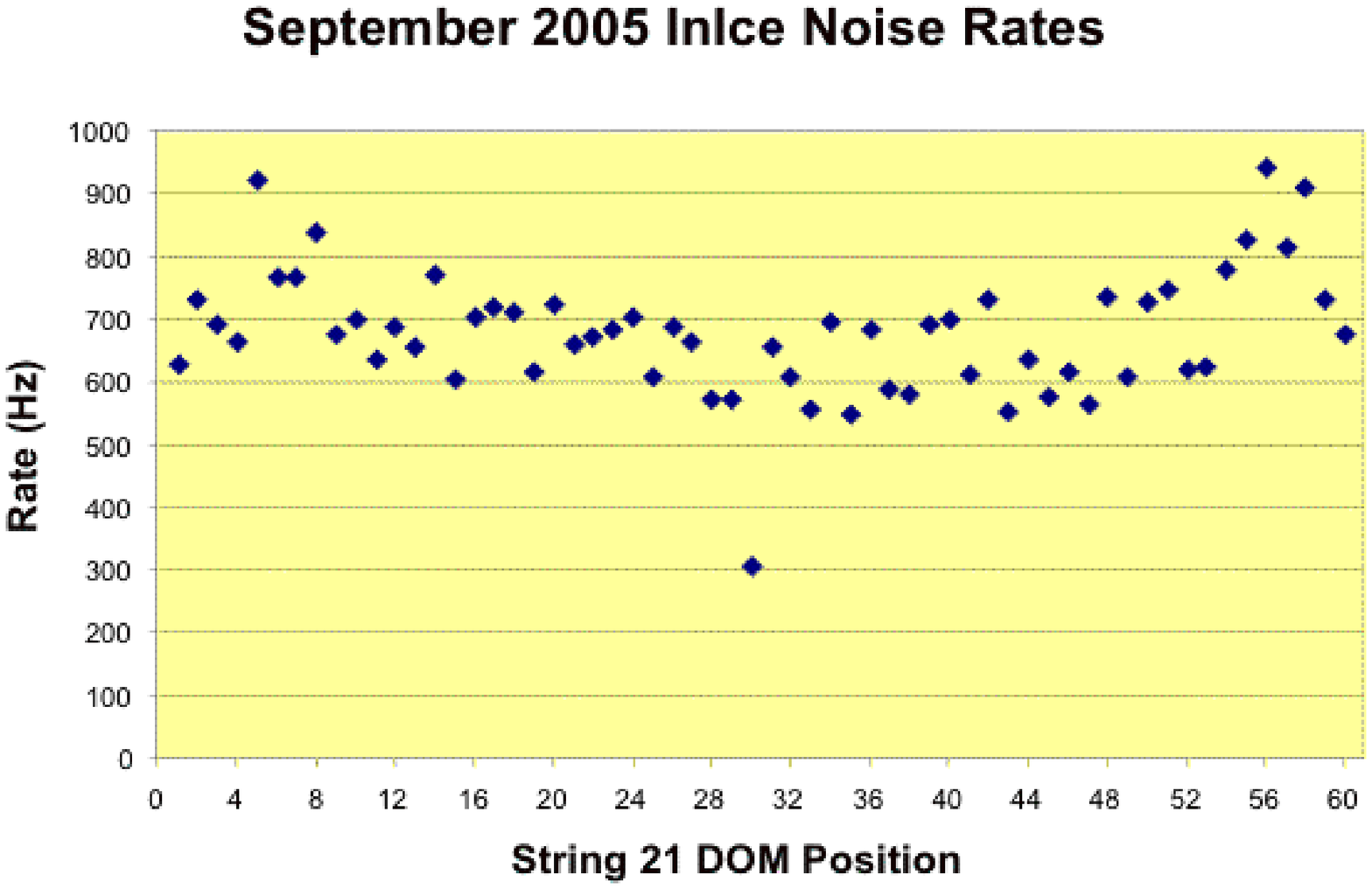}
\includegraphics{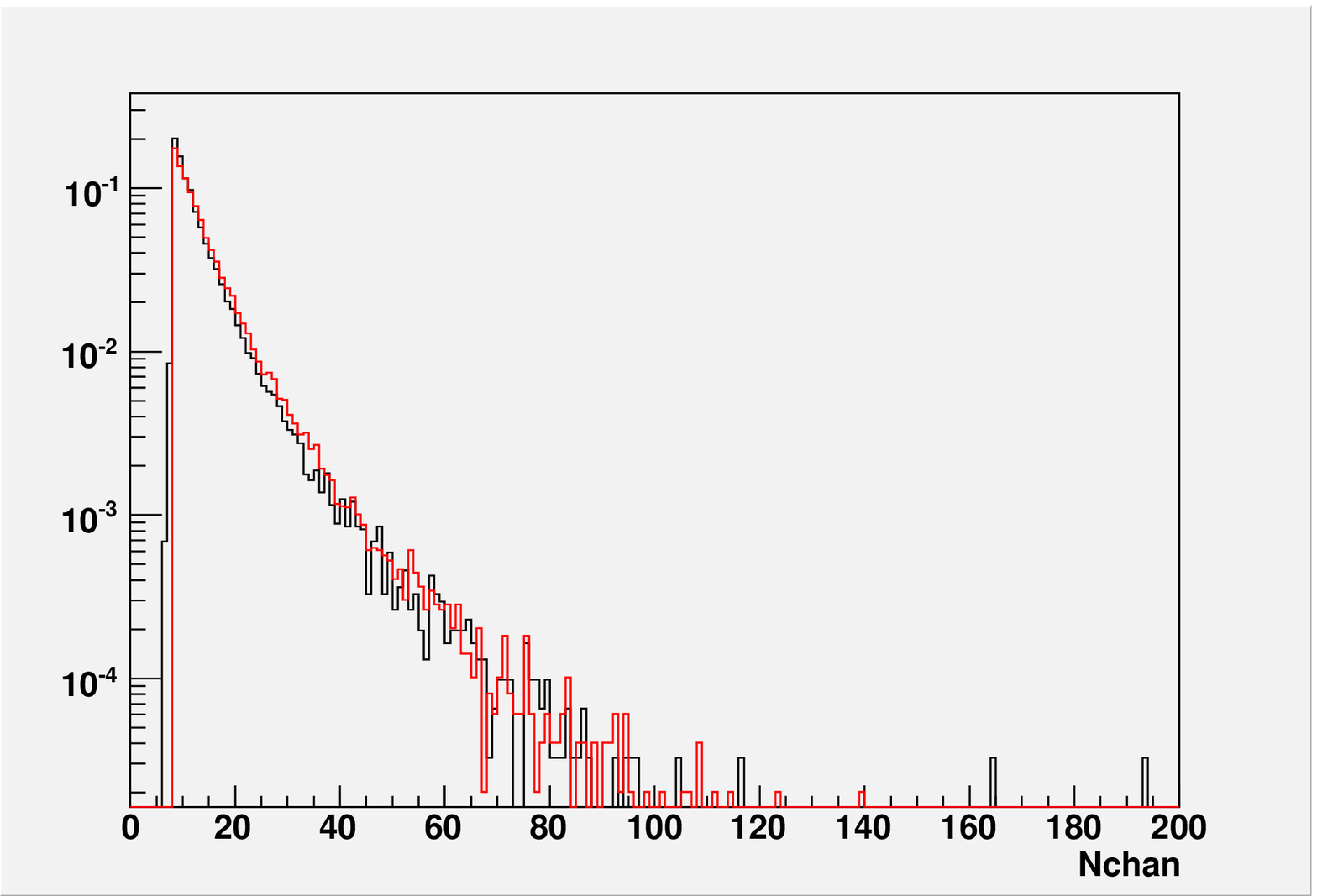}
      \caption[h]{{\bf On the left:} Noise rates versus DOM location of the first IceCube string.
      {\bf On the right:} Distribution of the number of hit DOMs in each event in a run of the 9 strings detector compared to the simulation of atmospheric muons (normalized to the number of events).}
     \label{fig3}
    \end{figure}

The main component of the IceCube detector is the DOM, an autonomous data collection unit consisting of a 25~cm Hamamatsu PMT and fast digitization electronics measuring the arrival time of every photon. Signals are processed by the Analog Transient Waveform Digitizer (ATWD) digitizing at 300~MHz for the first 400~ns and a Fast Analogue-to-Digital Converter (FADC) recording at 40~MHz for 6.4 $\mu$s.
The dynamic range is 500 photoelectrons (PE)/15~ns and 25000 PE/6.4 $\mu$s.
The dark noise rate in the ice has been measured for String 21 and it is shown in Fig.~\ref{fig3}(on the left): such low value allows for SN collapse searches which rely on increased counting rates over a few seconds. Moreover, the average dark noise rate
is reduced at the level of 400 Hz with a 51 $\mu$s deadtime suppression of afterpulses enhancing sensitivity for this search.
String 21 results, collected in (Ackermann et al., 2006b), comprise the reconstructed angular distribution
of muons in good agreement with simulations. Vertical up-going neutrino events have been observed, consistent with expectations of atmospheric neutrinos.
We are now working on the 9 string data and refining the simulation. At high energy the 9 string configuration is already larger than AMANDA-II. A preliminary result
that indicates we are beginning to understand the detector and its simulation is shown in 
Fig.~\ref{fig3}(on the right).
    
\section{Conclusions}

After 2 deployment seasons, IceCube is currently taking data in the configuration of 9 strings and 16 surface stations. Based on last season experience, the Collaboration expects to deploy
up to 14 strings per season and complete IceCube in the 2010-11 season.
Software development and physics analyses are ongoing and we expect first results very soon.

\end{document}